\documentclass[aps,prl,showpacs,twocolumn,groupedaddress]{revtex4}

\usepackage{amsmath,amssymb}
\usepackage{graphicx}
\usepackage{psfrag}

\begin{document}

\title{Statistical mechanics of Semiflexible Bundles of Wormlike Chains}

\author{Claus Heussinger}\author{Mark Bathe}\author{Erwin Frey}

\affiliation{Arnold Sommerfeld Center for Theoretical Physics and
  CeNS, Department of Physics, Ludwig-Maximilians-Universit\"at
  M\"unchen, Theresienstrasse 37, D-80333 M\"unchen, Germany}

\begin{abstract}
  We demonstrate that a semiflexible bundle of wormlike chains exhibits a
  state-dependent bending stiffness that alters fundamentally its scaling
  behavior with respect to the standard wormlike chain. We explore the
  equilibrium conformational and mechanical behavior of wormlike bundles in
  isolation, in crosslinked networks, and in solution.
\end{abstract}

 \date{\today}

\maketitle

In recent decades, the wormlike chain (WLC) has emerged as the standard model
for the description of semiflexible polymers~\cite{saito67}. The
defining property of a WLC is a mechanical bending stiffness, $\kappa_f$, that
is an intrinsic material constant of the polymer. Within this framework,
numerous correlation and response functions have been calculated, providing a
comprehensive picture of the equilibrium and dynamical properties of
WLCs~\cite{granek97,marko95,wilhelm96}. A number of experimental studies have
demonstrated the applicability of the WLC model to DNA~\cite{bustamante94} and
F-actin~\cite{legoff02}, among other biological and synthetic polymers.
Significant progress has also been made towards the description of the
collective properties of WLCs, for example, in the form of entangled solutions.
One of the hallmarks of this development is the scaling of the plateau shear
modulus with concentration, $G\sim c^{7/5}$~\cite{isa96,odj83,sem86}, which is
well established experimentally~\cite{hin98,xu98}.

Another important emerging class of semiflexible polymers consists of
\emph{bundles} of WLCs~\cite{claessens06NatMat,kis04}. Semiflexible polymer
bundles consisting of F-actin or microtubules are ubiquitous in
biology~\cite{lodish03}, and have unique mechanical properties that may well be
exploited in the design of nanomaterials~\cite{kis04}. As shown by Bathe {\it et
  al.}~\cite{bathe06,bathe07} wormlike bundles (WLB) have a state-dependent
bending stiffness, $\kappa_B$, that derives from a generic interplay between the
high stiffness of individual filaments and their rather soft relative
sliding motion. In this Letter, we demonstrate that this state-dependence gives
rise to fundamentally new behavior that cannot be reproduced trivially using
existing relations for WLCs.  We explore the consequences of a state-dependent
bending stiffness on the statistical mechanics of isolated WLBs, as well as on
the scaling behavior of their entangled solutions and crosslinked networks.

We consider the bending of ordered bundles with isotropic cross-section. A
bundle consists of $N$ filaments of length $L$ and bending stiffness $\kappa_f$.
Filaments are irreversibly crosslinked to their nearest neighbors by discrete
crosslinks with mean axial spacing $\delta$. Crosslinks are modeled to be
compliant in shear along the bundle axis with finite shear stiffness $k_\times$,
and to be inextensible transverse to the bundle axis, thus fixing the
interfilament distance, $b$~\footnote{Typical protein
  elongational stiffnesses are of the order $1 \rm{N/m}$ while they are much softer in
  shear~\cite{claessens06NatMat}.}.  Bundle deformations are characterized by
the transverse deflection $\mathbf{r}_\perp(s)$ of the bundle neutral surface at
axial position $s$ along the backbone and by the stretching deformation $u_i(s)$
of filament $i$. The torsional stiffness of the bundle is assumed to be of the
same order as the bending stiffness. Thus, as long as transverse deflections
remain small (``weakly bending'') the two components of $\mathbf{r}_\perp$ are
decoupled and effects of twist are of higher order~\cite{landau7}. The bundle
response may then be analyzed in planar deformation, where the bending stiffness
results from the superposition of $2M = \sqrt N$ bundle layers.

The WLB Hamiltonian consists of three contributions, $H_{\rm WLB}=H_{\rm
  bend}+H_{\rm stretch}+H_{\rm shear}$. The first term corresponds to the
standard WLC Hamiltonian
\begin{equation}\label{eq:hbend}
  H_{\rm bend} = \frac{N\kappa_f}{2}\int_0^L\!ds
  \left(\frac{\partial^2r_\perp}{\partial s^2}\right)^2\,, 
\end{equation}
which is the same for each of the $N$ filaments. The second term accounts for
filament stretching,
\begin{equation}\label{eq:hstretch}
  H_{\rm stretch} = Mk_s\delta\int_0^L\!ds \sum_{i=-M}^{M-1}
  \left(\frac{\partial u_i}{\partial s}\right)^2 \,,
\end{equation}
where $k_s$ is the single filament stretching stiffness on the scale of the
crosslink spacing $\delta$.  {No particular form for bending and stretching
  stiffnesses is assumed, but one may think of the filaments as homogeneous
  elastic beams with Youngs modulus $E$, for which $\kappa_f\sim Eb^4$ and
  $k_s\sim Eb^2/\delta$.  Alternatively, $k_s$ may represent the entropic
  elasticity of a WLC, for which $k_s\sim \kappa_f^2/T\delta^4$.}

\begin{figure}[t]
 \begin{center}
   \includegraphics[width=0.9\columnwidth]{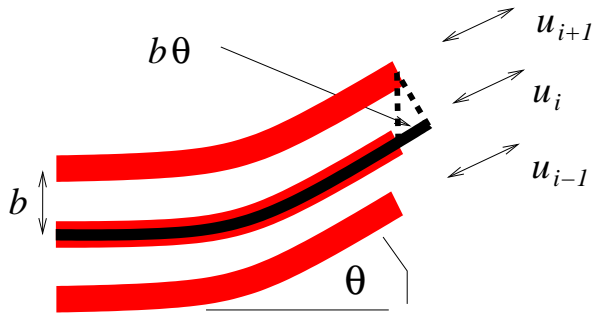}
\end{center}
\caption{Illustration of the geometry of a single bundle layer (The full bundle
  consists of $2M$ layers that are stacked in parallel.). The bundle is
  deflected through the angle $\theta=r_\perp'$. If filament $i$ stretches the
  amount $u_i=u_{i+1}+b\theta$, the crosslink (dashed line) remains undeformed
  with zero shear energy.}
  \label{fig:bundel_figure}
\end{figure}

The third energy contribution, $H_{\rm shear}$, results from the
crosslink-induced coupling of neighboring filaments. To minimize the crosslink
energy, any relative filament slip induced by cross-sectional rotations $\theta
= \partial_sr_\perp\equiv r_\perp'$ must be compensated by filament stretching
(Fig.~\ref{fig:bundel_figure}). This crosslink shear energy, which simply
suppresses relative sliding motion of neighboring filaments, is given by
\begin{equation}\label{eq:hshear}
H_{\rm shear} = \frac{Mk_\times}{\delta}\int_0^L\!ds
    \sum_{i=-M+1}^{M-1}(\Delta u_i+b\frac{\partial r_\perp}{\partial s})^2 \,,
\end{equation}
where $\Delta u_i = u_i - u_{i-1}$.
A related model for two filaments was introduced by Everaers {\it et al.}  in
Ref.~\cite{everaers95}, where special emphasis was placed on the limit of
inextensible filaments, $k_s\to\infty$. In that model, the anisotropic bundle
cross-section leads to a coupling of in-plane and out-of-plane bending
modes~\cite{mergell02} that is absent in the present model because it has a
symmetric cross-section.

Functional differentiation of the Hamiltonian results in the (overdamped)
equations of motion
\begin{eqnarray}\label{eq:variationRperp}
  N\kappa_fr_\perp'''' - \frac{2Mk_\times b}{\delta}\sum_{i}(\Delta u_i'+br_\perp'')
  & = & F(r_\perp,s) \,, \\ \label{eq:variationUk} 
  k_s\delta u_i''+\frac{k_\times}{\delta}(\Delta u_{i+1}-\Delta u_i) & = & 0\,,
\end{eqnarray}
where $F$ is a transverse force that may represent fluid drag, random thermal
noise, or other external loading. To proceed, Eq.~(\ref{eq:variationUk}) is
solved together with appropriate boundary conditions, so as to eliminate the
$u_i$ in Eq.~(\ref{eq:variationRperp}).  The calculations are most easily
performed in Fourier-space, where we write for the expansions $r_\perp(s)
=\sum_n r_n \sin(n\pi s/L)$ and $u_i(s) = \sum_n u_{in} \cos(n\pi s/L)$,
applicable to pinned boundary conditions. The resulting equation of motion for
$r_n$ then takes the simple form $\kappa_nq_n^4r_n = F_n$, with a mode-number
dependent effective bending stiffness $\kappa_n$. The general result for
$\kappa_n$ is obtained using the standard ansatz $u_i\sim w^i$, which reduces
Eq.~(\ref{eq:variationUk}) to an equation that is quadratic in $w$.

In the following, we present an approximate solution to
Eqs.~(\ref{eq:variationRperp}) and~(\ref{eq:variationUk}) that is based on the
assumption that filament stretching increases \emph{linearly} through the bundle
cross-section, $u_i = \Delta u\cdot(i+1/2)$~\cite{tolomeo97}.  Although
comparison with the exact solution demonstrates that $u_i$ in general varies
nonlinearly with $i$~\cite{tbpCH}, it turns out that the effective bending
stiffness $\kappa_n$ is insensitive to this nonlinearity. At the same time, the
linearization simplifies the formulas substantially, so that the effective
bending stiffness is given in closed form by
\begin{equation}\label{eq:kappaLINEAR}
  \kappa_n = N\kappa_f\left[ 1 + \left( \frac{12\hat\kappa_f}{N-1} +
      (q_n\lambda)^2   \right)^{-1} \right]\,,  
\end{equation}
with a dimensionless bending stiffness $\hat\kappa_f=\kappa_f/k_s\delta b^2$ and
a length-scale $\lambda = (L/\sqrt{\alpha})\sqrt{M\hat\kappa_f/(M-1/2)}$, that
depends on the shear stiffness $k_\times$ via the dimensionless coupling
parameter $\alpha=k_\times L^2/k_s\delta^2$.

For any given mode-number $q_n\sim n/L$, three different elastic regimes emerge
as asymptotic solutions for $N \gg 1$ and respective values of
$\alpha$~\cite{bathe06,bathe07}. For large shear stiffness ($\alpha\gg N$), the
\emph{fully coupled} bending scenario is obtained, where the bundle behaves like
a homogeneous beam with $\kappa_n\sim N^2k_s$. For intermediate values of the
shear stiffness ($1 \ll \alpha\ll N$), the bending stiffness in the \emph{shear
  dominated} regime is $\kappa_n\sim Nk_\times q_n^{-2}$ and the full
mode-number dependence of Eq.~(\ref{eq:kappaLINEAR}) has to be accounted for.
Finally, \emph{decoupled bending} of $N$ laterally independent, but transversly
constrained, filaments is found in the limit of small cross-link shear stiffness
($\alpha\ll 1$), where the bending stiffness is simply $\kappa_n=N\kappa_f$.

In the particular limit of $N\to\infty$ and fixed bundle diameter
$D=b\sqrt{N}\ll L$, Eq.~(\ref{eq:kappaLINEAR}) reduces to the Timoshenko model
for beam bending~\cite{timoshenko}, which was recently used to interpret bending
stiffness measurements on microtubules~\cite{kis02,pampaloni06} and carbon
nanotube bundles~\cite{kis04}. In this limit
\begin{equation}\label{eq:kappaTIMO}
  \kappa_n = \frac{N^2\kappa_f}{1+(q_nD)^2E/12G}\,,
\end{equation}
where we have used the expressions of $k_s$ and $\kappa_f$ for homogeneous beams
and defined $G=k_\times/\delta$. While this limit serves as a consistency check
for our mathematical analysis, real bundles consist of a finite, and often
small, number of constituent filaments, for which Eq.~(\ref{eq:kappaTIMO})
cannot be applied to describe the full range of bending behavior captured by
Eq.~(\ref{eq:kappaLINEAR}). Indeed, in Eq.~(\ref{eq:kappaTIMO}) no decoupled
bending regime exists and the bending stiffness vanishes as the crosslink shear
stiffness approaches zero~\footnote{Re-analyzing the data of~\cite{pampaloni06}
  using Eq.~(\ref{eq:kappaLINEAR}) we expect microtubules shorter than $L\approx
  3.5\mu{\rm m}$ to be in the decoupled regime with a constant persistence
  length, $l_p\approx 0.18{\rm mm}$.}. The condition, $\alpha\sim N$,
delineating the remaining two regimes can be rewritten as $E/G\sim (L/D)^2\gg
1$, which re-emphasizes the small value of cross-link shear stiffness in the
intermediate regime.

For fixed values of $(N,\alpha)$, the bundle bending stiffness
Eq.~(\ref{eq:kappaLINEAR}) crosses over from fully coupled to decoupled bending
via the intermediate regime as the mode-number $q_n$ is increased. Thus,
different modes may belong to different elastic regimes, rendering the
fluctuation properties of the bundle non-trivial and qualitatively different
from single semiflexible polymers. This cross-over is mediated by the
length-scale $\lambda$, which acts as a cut-off on the fluctuation spectrum:
whereas wavelengths $q_n^{-1} \ll \lambda$ belonging to the decoupled regime are
characterized by a constant bending stiffness, modes with $q_n^{-1} \gg \lambda$
acquire a higher stiffness $\kappa_n\sim q_n^{-2}$ and are thereby suppressed.
Finally, for even longer wavelengths $q_n^{-1}\gg \lambda \sqrt{N}$, the bending
stiffness reattains a constant, limiting value.  As an example (taken from
Ref.~\cite{claessens06NatMat}) we found $\lambda\approx 7\mu m$ for actin/fascin
bundles with $N\approx 30,\,L\approx50\mu m$.

In situations where modes pertaining to the intermediate regime are irrelevant,
the $q$-dependence of $\kappa_n$ drops out and one recovers the single WLC
result, albeit with a renormalized persistence length $l_p\to N l_p$ in the
decoupled, and $l_p\to N^2 l_p$ in the fully coupled, regimes, respectively.  In
other cases, calculation of the tangent-tangent correlation function
demonstrates that the persistence length cannot be defined unambiguously. As
indicated in Ref.~\cite{everaers95}, the correlation function does not decay
exponentially, but rather exhibits a complex structure at intermediate
distances~\cite{tbpCH}. In the following, we will therefore explore the
consequences on the statistical mechnanics of the WLB in particular as regards
the intermediate regime.

\begin{figure}[t]
 \begin{center}
   \includegraphics[width=0.85\columnwidth]{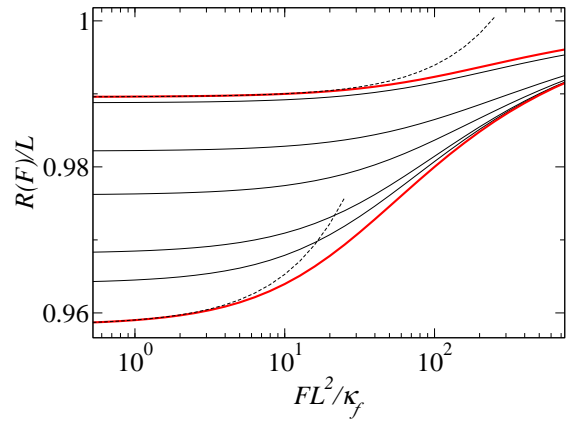}
\end{center}
\caption{End-to-end distance $R(F)/L$ as a function of stretching force $F
  L^2/\kappa_f$ for a bundle of $N=4$ filaments and $L=l_p$. The black curves
  correspond to $\lambda/L=0.01,0.1,..,0.7$. Thick red curves relate to (bottom)
  decoupled and (top) fully coupled bending, respectively. Dashed lines
  correspond to the respective linear response regimes.}
  \label{fig:force_extension}
\end{figure}

First, consider the force-extension relation as calculated from the end-to-end
distance $R(F)=L - \sum_nk_BT/(\kappa_n q_n^2+F)$, where $F$ is the force
applied to the bundle ends~\footnote{Note, that a second contribution from the
  extensibility of the bundle backbone is neglected here, for simplicity.}.  For
small stretching forces one may readily calculate the linear response
coefficient $k_{\rm entr} = F/(R(F) - R(0))$ using a Taylor series expansion.
The result in the intermediate regime is
\begin{equation}\label{eq:}
  k_{\rm entr} \propto \frac{(N\kappa_f)^2}{L\lambda^3k_BT}\,, \quad (\lambda
  \sqrt{N} \gg L \gg \lambda)\,. 
\end{equation}
which is inversely proportional to bundle length, like a mechanical beam.
Importantly, the strong dependence of $k_{\rm entr}(L)\sim L^{-4}$ applicable to
single filaments (and the other two regimes) is lost. This has dramatic
consequences on the plateau value of the shear modulus $G$ in crosslinked bundle
networks, which in affine theories~\cite{rubinstein03} is assumed to be given in
terms of $k_{\rm entr}$ by $G\sim k_{\rm entr}(\xi)/\xi$, where the mesh-size
$\xi$ depends on concentration $c$ as $\xi\sim c^{-1/2}$. Accordingly, in the
intermediate regime one finds $G\sim c$, which is a much weaker
concentration-dependence than $G\sim c^{5/2}$~\cite{mac95} applicable to single
filaments. It is worthwhile noting that the force-extension relation is strongly
nonlinear (see Fig.\ref{fig:force_extension}), rendering the linear response
valid only for very small relative extensions. In this particular example the
linear response formula deviates from the exact solution by $50\%$ at only
$\approx 3\%$ and $\approx 0.7\%$ strain in the decoupled and the fully coupled
limits, respectively.

Bundle behaviour under compressive forces further highlights the unusual
properties of WLBs. Because the bending stiffness in the intermediate regime
scales with the length of the bundle as $\kappa_B\sim L^2$, the Euler buckling
force $F_c\sim \kappa_B/L^2\sim N\kappa_f/\lambda^2$ is \emph{independent} of
bundle length. This unique property may well be exploited in polymerizing
biological bundles such as filopodia, which may increase their contour length
against compressive loads without loss of mechanical stability.

\begin{figure}[t]
 \begin{center}
   \includegraphics[width=0.85\columnwidth]{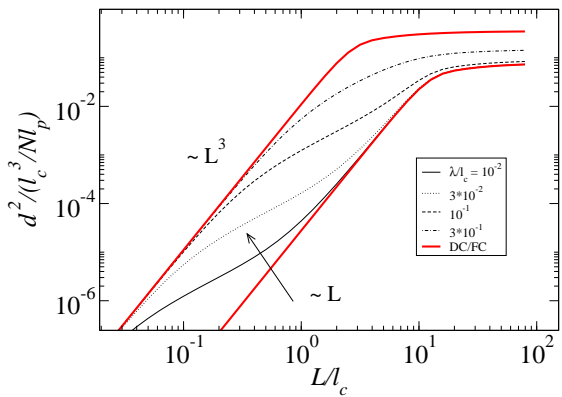}
\end{center}
\caption{Tube diameter $d^2/(l_c^3/Nl_p)$ as a function of contour length
  $L/l_c$ for various $\lambda/l_c$ and $M=20$. Thick (red) curves correspond to
  (top) decoupled and (bottom) fully coupled bending, respectively. For short
  filaments the intermediate regime is visible through the linear slope $d^2\sim
  L$ (see Eq.~(\ref{eq:transverse-fluctuations})). For long filaments the
  fluctuations saturate.  By increasing $\lambda$ the tube is becoming wider
  (Eq.~(\ref{eq:tubeDiameter})).}
  \label{fig:tube_diameter}
\end{figure}

Complementary to the elasticity of crosslinked networks of WLBs, we turn next to
the elasticity of their entangled solutions. The generally accepted theory for
the concentration dependence of the plateau modulus of entangled WLCs is based
on the free energy change $\Delta F$ of confining a polymer to a tube of
diameter $d$~\cite{isa96,odj83}. The associated change in free energy is written
as $\Delta F \sim k_BTL/l_d$, which defines the deflection length $l_d$ to be
the scale at which the polymer starts to interact with its enclosing tube. The
deflection length itself is connected to the tube diameter $d$ and the filament
concentration $c$ via the standard excluded volume argument~\cite{sem86},
$l_d^2d = l_d/cL$, which balances the excluded volume of the tube with the
available volume per filament.  All that remains is the calculation of the tube
diameter $d$ of a single polymer confined by the potential
\begin{equation}\label{eq:hconfining}
  V_{\rm } = \frac{N\kappa_f}{2l_c^4}\int_0^L\!ds\, r_\perp^2(s)\,,
\end{equation}
where the confinement length $l_c$ is defined as a measure of the strength of
the potential. While $l_c\equiv l_d$ in the standard WLC, we will see shortly
that this does not hold for WLBs in the intermediate regime. First, consider the
transverse fluctuations of an unconfined bundle, in particular the average value
$d_0^2 \equiv \frac{1}{L}\int_s\langle r_\perp(s)^2 \rangle$. This is most
easily calculated as
\begin{equation}\label{eq:transverse-fluctuations}
  d_0^2 \sim    L\lambda^2/Nl_p\,,
  \quad (\lambda \sqrt{N} \gg L \gg \lambda)
\end{equation}
which has to be compared to the WLC result for which $d_0^2 \sim L^3/l_p$. In
the presence of the confining potential, the same calculation yields
\begin{equation}\label{eq:tubeDiameter}
  d^2 \sim     l_c^2\lambda/Nl_p\,, \quad (\lambda \sqrt{N} \gg l_c \gg \lambda)\,.
\end{equation}
For strong confinement $l_c\ll \lambda$, the potential suppresses all modes of
the intermediate regime and one recovers the expression valid for single
filaments, $d^2\sim l_c^3/l_p$. The general result for the tube diameter is
depicted in Fig.~\ref{fig:tube_diameter}.  As the contour length $L$ of the
bundle is increased it begins to ``feel'' the presence of its enclosing tube at
the deflection length $L=l_d$.  By comparing
Eq.~(\ref{eq:transverse-fluctuations}) with Eq.~(\ref{eq:tubeDiameter}) one
finds $l_d\sim l_c^2/\lambda$, which is valid in the intermediate regime. At the
same time, $l_d\equiv l_c$ in the decoupled and fully coupled regimes, where the
deflection and confinement lengths are identical.

One may use these results to rewrite the deflection length as a function of
concentration $c$.  In the intermediate regime the result is $l_d^3 \sim
Nl_p/(\lambda cL)^2$, which replaces the usual result $l_d^5\sim Nl_p/(cL)^2$
valid in the decoupled regime (strong confinement). The free energy of
confinement and the elastic plateau modulus $G\sim (cL)\Delta F/L$ now depend on
$\lambda$ and thus on the properties and density of the crosslinks.  The modulus
displays a cross-over that is \emph{mediated by concentration},
\begin{equation}\label{eq:freeEnergyC} 
   G \sim k_BT 
  \begin{cases} 
    (cL)^{5/3} (Nl_p)^{-1/3}\lambda^{2/3}\,, & \text{$c \ll c^\star$}, \\
    (cL)^{7/5}(Nl_p)^{-1/5}\,, & \text{$c \gg c^\star$},
  \end{cases}
\end{equation} 
where we defined the cross-over concentration as $(cL)^\star \sim \sqrt{Nl_p}
\lambda^{-5/2}$. Below the even smaller concentration $c^{\star\star}\sim
c^\star N^{-3/4}$, the fully coupled regime is entered and the modulus again
scales as $G\sim c^{7/5}$.

Having addressed equilibrium properties of WLBs, further consequences of the
state-dependent bending stiffness on dynamic response functions remain to be
explored, along with the effects of nonpermanent crosslinks.  Additional
experiments~\cite{claessens06NatMat,lieleg,tolomeo97,kis04} are required to test
the applicability of the derived results to biological and synthetic bundles.

\acknowledgments

Funding from the Alexander von Humboldt Foundation (to MB), from the German
Science Foundation (SFB486), and from the German Excellence Initiative via the
program "Nanosystems Initiative Munich (NIM)" is gratefully acknowledged.


\end{document}